\documentclass[letter]{aa}
\usepackage{graphicx}
\usepackage{txfonts}

\begin{document}

\title{Detection of a 1.59\,h period in the B supergiant star HD\,202850\thanks{Based on observations performed with the 2m-telescope at Ond\v{r}ejov Observatory}}

\author{M. Kraus\inst{1}, S. Tomi\'{c}\inst{2}, M.E. Oksala\inst{1}
\and
 M. Smole\inst{2}
  }
                                                                                
\institute{Astronomick\'y \'ustav, Akademie v\v{e}d \v{C}esk\'e republiky, 
Fri\v{c}ova 298, 251\,65 Ond\v{r}ejov, Czech Republic\\
\email{kraus@sunstel.asu.cas.cz} 
\and
Department of Astronomy, Faculty of Mathematics, University of Belgrade, 
Studentski trg 16, 11000 Belgrade, Serbia
      }
                                                                                
\date{Received; accepted}

\abstract
{Photospheric lines of B-type supergiants show variability in their profile 
shapes. In addition, their widths are much wider than can be accounted for 
purely by stellar 
rotation. This excess broadening is often referred to as macroturbulence. Both 
effects have been linked to stellar oscillations, but B supergiants have 
not been systematically searched yet for the presence of especially short-term 
variability caused by stellar pulsations.}
{We have started an observational campaign to investigate the line profile 
variability of photospheric lines in a sample of Galactic B supergiants. These 
observations aim to improve our understanding of the physical effects 
acting in the atmospheres of evolved massive stars.}
{We obtained four time-series of high-quality optical 
spectra for the Galactic B supergiant HD\,202850. The spectral coverage of 
about 500\,\AA \ around H$\alpha$ encompasses the \ion{Si}{ii}\,$\lambda\lambda 
6347, 6371$, and the \ion{He}{i}\,$\lambda 6678$ photospheric lines. 
The line profiles were analysed by means of the moment method.}
{The time-series of the photospheric \ion{Si}{ii} and \ion{He}{i} lines display 
a simultaneous, periodic variability in their profile shapes. Proper analysis 
revealed a period of 1.59\,h in all three lines. This
period is found to be stable with time over the observed span of 19 months.
This period is much shorter than the rotation period of the star and might be 
ascribed to stellar oscillations. Since the star seems to fall outside the 
currently known pulsational instability domains, the nature of the discovered 
oscillation remains unclear.}
{}


\maketitle

\section{Introduction}

Many B-type supergiants are reported to show spectroscopic and
photometric variability (e.g., Fullerton et al. \cite{Fullerton};
Kaufer et al. \cite{Kaufer97,Kaufer06}; Lefever et al. \cite{Lefever};
Markova et al. \cite{Markova}; Clark et al. \cite{Clark}). In addition,
spectral lines of B supergiants are found to be much wider
than expected from pure stellar rotation. This extra 
broadening, which is generally referred to as macroturbulence, can be
of the same order as the stellar rotation (e.g., Ryans 
et al. \cite{Ryans}; Sim\'{o}n-D\'{i}az \& Herrero \cite{SDH}; Markova \& 
Puls \cite{MarkovaPuls}). The nature of macroturbulence has long been
unclear, although the often high, supersonic value 
makes a kinematic origin of this broadening rather doubtful.

Lucy (\cite{Lucy}) previously suggested that the appearance of this
macroturbulence together with spectroscopic and photometric
variability could be a result of stellar pulsations. Recently, Aerts et 
al.\,(\cite{Aerts2009}) studied the shape of line profiles resulting 
from a combination of both non-radial pulsations and stellar rotation.
They found that numerous superimposed pulsational modes can result in
a substantial line width comparable to the rotational one. Observational 
evidence for the presence of non-radial pulsations in some B supergiants 
was found from spectroscopy as well as from space-based photometry  
(Saio et al. \cite{Saio}; Lefever et al. \cite{Lefever}). 
Sim\'{o}n-D\'{i}az et al. (\cite{Simon-Diaz2010})
found additional evidence for a correlation between macroturbulent broadening
and line profile variabilities.

The simultaneous presence of macroturbulent broadening and line profile 
variation in B supergiants hence seems to be directly linked to non-radial 
pulsations. Most periods identified in B supergiants so far are on the order 
of days and could be interpreted as classical gravity modes. 
Interestingly, in the analysis of photometric data for the late-type B 
supergiant HD\,46769 Lefever et al. (\cite{Lefever}) found a period of 
2.693\,h. In addition, the location of this object in the Hertzsprung-Russell 
diagram falls outside the instability domain calculated by Saio et al. 
(\cite{Saio}) for evolved massive stars that undergo gravity-mode pulsations. 
Here, we report on the discovery of another late-type B supergiant, HD\,202850, 
that seems to be located slightly outside this instability domain, but with a 
short-period variability. 
HD\,202850 (= $\sigma$\,Cyg) is a Galactic supergiant classified as
B9\,Iab. It is located in the OB association Cyg\,OB\,4 at a distance of
$\sim 1$\,kpc (Humphreys \cite{Humphreys}). The set of stellar parameters
obtained by Markova \& Puls (\cite{MarkovaPuls}) is given in
Table\,\ref{Tab:param}.

\begin{figure*}[t!]
\resizebox{\hsize}{!}{\includegraphics{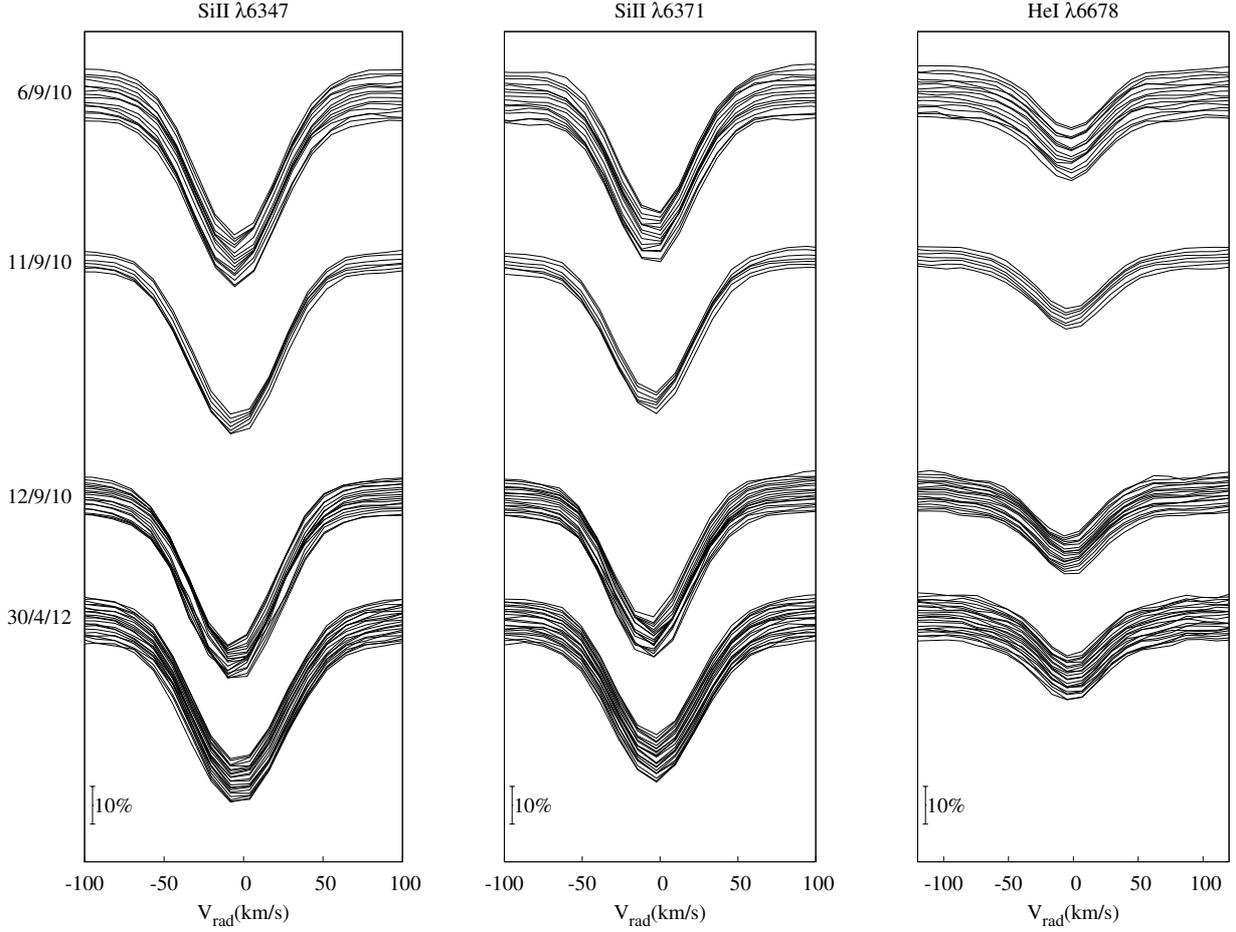}}
\caption{Time-series of the \ion{Si}{ii}\,$\lambda\lambda 6347, 6371$ and 
\ion{He}{i}\,$\lambda 6678$ lines. The time increases from top to bottom. 
For better visual inspection consecutive spectra have been slightly shifted 
along the vertical axis with offsets proportional to the exposure times.}
\label{Fig:timeseries}
\end{figure*}

\section{Observation and data reduction}\label{obs}

\begin{table}
\caption{Stellar parameters of HD\,202850 from Markova \& Puls (\cite{MarkovaPuls}).}
\label{Tab:param}
\centering
  \begin{tabular}{c c c c c c c}
    \hline
    \hline
$T_{\rm eff}$ & $\log L/L_{\odot}$ & $\log g$ & $R_{*}$ & $M$  & $\varv\sin i$ & $\varv_{\rm macro}$ \\
$ $ [K]          &               &   [cgs] & [$R_{\odot}$] & [$M_{\odot}$] & [km\,s$^{-1}$] & [km\,s$^{-1}$] \\
   \hline
11\,000       &  4.59         &   1.87   &   54          & $8^{+4}_{-3}$ & $33\pm 2$ & $33\pm 2$  \\
    \hline
  \end{tabular}
\end{table}

\begin{table}
\caption{Journal of observations. The Heliocentric Julian Date (HJD) refers 
to the middle of the exposure.}
\label{Tab:obs}
\centering
\begin{tabular}{cccccc}
    \hline
    \hline
HJD  & $t_{\rm exp}$ & HJD  & $t_{\rm exp}$ & HJD  & $t_{\rm exp}$ \\
($2450000+$) & [s]  & ($2450000+$) & [s]  & ($2450000+$) & [s]  \\ 
\hline
5466.37462 &       600     &       5451.53004 &       600     &       6048.50304       &       300     \\
\cline{3-4}
5466.38362 &       600     &       5452.32982 &       250     &       6048.50843       &       300     \\
5466.39268 &       600     &       5452.33502 &       300     &       6048.51384       &       300     \\
5466.40173 &       600     &       5452.34064 &       300     &       6048.51925       &       300     \\
5466.41073 &       600     &       5452.34610 &       300     &       6048.52465       &       300     \\
5466.41974 &       600     &       5452.35213 &       300     &       6048.53005       &       300     \\
5466.42877 &       600     &       5452.35767 &       300     &       6048.53547       &       300     \\
5466.43781 &       600     &       5452.36317 &       300     &       6048.54087       &       300     \\
5466.44683 &       600     &       5452.36868 &       300     &       6048.54629       &       300     \\
5466.45582 &       600     &       5452.37426 &       300     &       6048.55170       &       300     \\
5466.46486 &       600     &       5452.37977 &       300     &       6048.55710       &       300     \\
5466.47388 &       600     &       5452.38528 &       300     &       6048.56252       &       300     \\
5466.48296 &       600     &       5452.39076 &       300     &       6048.56794       &       300     \\
5466.49201 &       600     &       5452.39631 &       300     &       6048.57336       &       300     \\
5466.50103 &       600     &       5452.40179 &       300     &       6048.57877       &       300     \\
5466.51002 &       600     &       5452.40736 &       300     &       6048.58416       &       300     \\
\cline{1-2}
5451.47616 &       600     &       5452.41293 &       300     &       6048.58957       &       300     \\
5451.48516 &       600     &       5452.41850 &       300     &       6048.59496       &       300     \\
5451.49411 &       600     &       5452.42410 &       300     &       6048.60036       &       300     \\
5451.50312 &       600     &       5452.42976 &       300     &       6048.60576       &       300     \\
\cline{3-4}
5451.51209 &       600     &       6048.49219       &       300     &       6048.61125       &       300     \\
5451.52107 &       600     &       6048.49762       &       300     &       & \\
\hline
\end{tabular}
\end{table}

We observed HD\,202850 on 2010 September 6, 11, and 12 and on 2012 April
30 (see Table\,\ref{Tab:obs}),
using the Coud\'{e} spectrograph attached to the 2-m telescope at Ond\v{r}ejov
Observatory (\v{S}lechta \& \v{S}koda \cite{Mirek}). We used the 830.77 lines
mm$^{-1}$ grating with a SITe $2030\times 800$ CCD that delivered a spectral
resolution of $R\simeq 13\,000$ in the
H$\alpha$ region with a wavelength coverage from 6253\,\AA \ to  6764\,\AA.
For wavelength calibration, a comparison spectrum of a ThAr lamp was taken
immediately after each exposure. The stability of the wavelength scale was
verified by measuring the wavelength centroids of O{\sc i} sky lines. The
velocity scale remained stable within 1\,km\,s$^{-1}$.

The data were reduced and heliocentric velocity corrected using standard
IRAF\footnote{IRAF is distributed by the National Optical Astronomy
Observatories, which are operated by the Association of Universities for
Research in Astronomy, Inc., under cooperative agreement with the National
Science Foundation.} tasks. On each night we also observed a rapidly rotating
star (HR\,7880, Regulus) to perform the telluric correction. Final
ranges in signal-to-noise ratios are 250--500, and the data with the highest
quality were those obtained on 2010 September 12.

\section{Results}\label{results}

\begin{figure*}[t!]
\resizebox{\hsize}{!}{\includegraphics{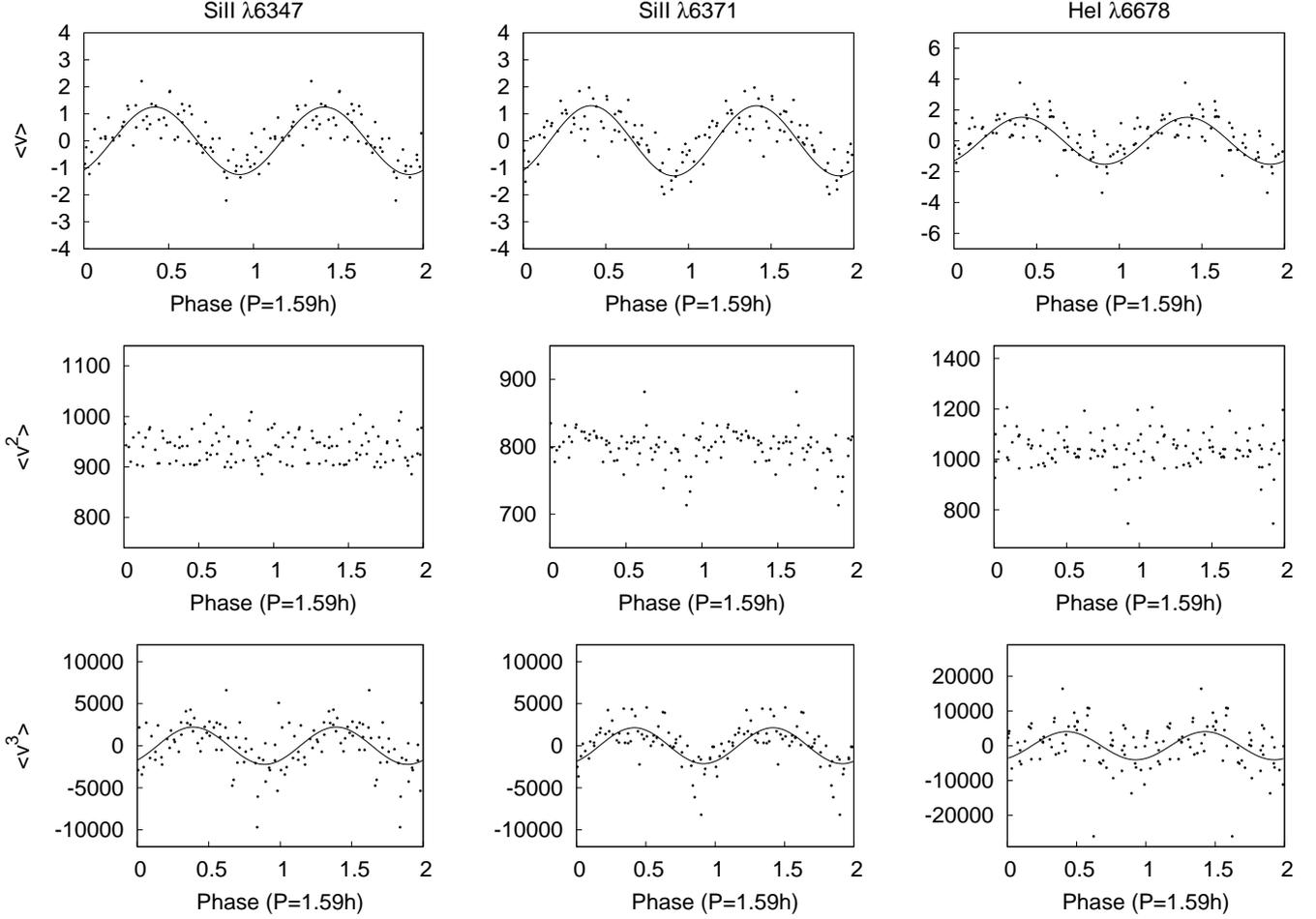}}
\caption{First three observed moments of the \ion{Si}{ii}\,$\lambda\lambda 
6347, 6371$ and \ion{He}{i}\,$\lambda 6678$ lines (dots). $<\varv>$, 
$<\varv^{2}>$, and $<\varv^{3}>$, have units km\,s$^{-1}$, (km\,s$^{-1}$)$^{2}$,
and (km\,s$^{-1}$)$^{3}$, respectively. The solid lines represent the best-fit 
sine curves to the first and third moments.} 
\label{Fig:moments}
\end{figure*}

In addition to H$\alpha$, our observed spectral range also included the 
photospheric lines of \ion{Si}{ii}\,$\lambda\lambda 6347, 6371$, and 
\ion{He}{i}\,$\lambda 6678$. Their time-series obtained on the four 
nights are shown in
Fig.\,\ref{Fig:timeseries}. A periodic variation is visible simply
from visual inspection, and the behaviour of the variation is similar in 
all the lines. If the variation is indeed periodic, the period must be 
shorter than (or similar to) the time span of the observations, which were 
$\sim 3.5, \sim 1.5, \sim 2.5$, and $\sim 3.0$\,h.
  
As mentioned above, line profile variability in B supergiants might be caused
by non-radial pulsations. Until now, the associated periods were on the order 
of days. To check whether pulsations might cause these short-term variations 
in the line profiles of HD\,202850, and to quantify this possible period, we 
applied the moment method, following the description of Aerts et al. 
(\cite{Aerts1992}) and North \& Paltani (\cite{NorthPaltani}). These moments 
have proven to be excellent tools for studying line profile variabilities 
because their behaviour allows one to distinguish between different origins of
the variability, such as stellar spots, radial and non-radial pulsations.    
We are aware that the quality of our data, especially the resolution of
our spectrograph with only $R \simeq 13\,000$, is not sufficient to perform
a successful mode identification, which would require a spectral resolution of
$R > 40\,000$ (Zima \cite{Zima}). Nevertheless, the moment method applied to 
our data will unveil a possible periodicity in the line profiles.

\begin{figure}[t!]
\resizebox{\hsize}{!}{\includegraphics{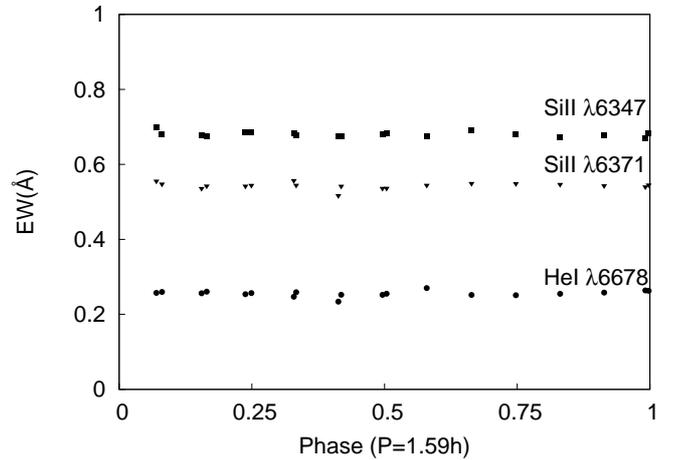}}
\caption{Stability in equivalent widths of the three photospheric lines.
Shown are the phased data from 2010 Sep. 12, which have the highest quality.}
\label{Fig:EW}
\end{figure}

For \ion{He}{i} and the \ion{Si}{ii} lines, the results of the moment 
calculations for the combined data set of all four time-series are 
displayed in Fig.\,\ref{Fig:moments}.
The moments are plotted versus phase, using the period obtained from 
two different methods applied to the first moments of the 
two nights with the shortest exposure times: (i) a simple sine curve fit and 
(ii) a more sophisticated Fourier analysis. The period obtained from all three 
lines applying both methods is $P=1.59\pm 0.01$\,h.
The moments obtained from the time-series in the other two nights were phased 
with the same period. Although these longer exposure observations suffer from 
some smearing effects, their moments still display the same periodic behaviour.

\section{Discussion}\label{disc}

The moment analysis delivers important insight into the possible 
nature of the line profile variability. The simultaneous variation in the 
line profiles of both helium and metal lines means that the presence of stellar 
spots can be excluded with a high probability. These spots are common in
chemically peculiar stars and are usually related to surface inhomogeneities in 
the form of over- and under-abundances in different elements. The photospheric
line profiles of these elements hence vary with the rotation period of the 
star, and the variations in the lines from different elements are typically not 
identical (e.g., Briquet et al. \cite{Briquet01, Briquet04}; Lehmann et al.
\cite{Lehmann}). In addition, stars with a patchy surface abundance pattern are 
known to show rotationally modulated light variability (Krti\v{c}ka et al. 
\cite{Jirka07, Jirka09}). From the Hipparcos light-curve of HD\,202850,
Koen \& Eyer (\cite{KoenEyer}) derived a putative (but not yet confirmed) 
period of 120.2\,d. This period might be caused by stellar rotation, but
the value of $\varv\sin i = 33\pm 2$\,km\,s$^{-1}$ obtained by Markova \& Puls 
(\cite{MarkovaPuls}) limits the rotation period to $P \le (83\pm 5)$\,d.

The presence of an equal contribution of macroturbulent and rotational
broadening in the line profiles of HD\,202850 (see Table\,\ref{Tab:param}) 
together with the simultaneous variation in the moments of both the 
\ion{He}{i} and the \ion{Si}{ii} lines seems therefore to speak in favour of 
pulsations as the origin of the line profile variations. 

The equivalent widths of the lines phased to the 1.59\,h period (Fig.\,\ref{Fig:EW})
do not show a periodic variability, but like most pulsating stars have variations
of a few percent at most (e.g., De Ridder et al. \cite{DeRidder}).
The simple sine curve fits to the third moments (Fig.\,\ref{Fig:moments}) imply
that these moments are dominated by the main frequency. The second moments
show a large scatter that makes it difficult to claim specific variations, although
indications for a possible double sine (corresponding to twice the main frequency)
seem to be present (especially for \ion{Si}{ii}\,$\lambda 6371$), 
which would speak in favour of an axisymmetric mode. 
However, these moments suffer
strongest from the noise introduced by the signal-to-noise ratio and the
relatively low resolution of the data, which severely hampers a proper analysis of
their real variations. Hence from the quality of our data we are only able to
confirm the presence of the period. For a proper mode identification, 
spectroscopic data with much higher resolution and signal-to-noise ratio are 
needed.

While periods on the order of hours are common
for non-radially pulsating stars e.g. on the main sequence (e.g. Smith
\cite{Smith}), pulsation periods in B supergiants have been reported to be
on the order of days from both observations and theory. Our
period of less than 2\,h therefore does not seem to fit into the currently known
domain of non-radially pulsating supergiants.

In a recent study, Lefever et al. (\cite{Lefever}) searched for periodicities
in the Hipparcos data of a sample of B supergiants. The shortest period they
found was 0.1122\,d (= 2.693\,h) for the B8\,Ib supergiant star HD\,46769.
While Lefever et al. (\cite{Lefever}) did not comment on the period found for 
this star in more detail, it is interesting to report here that our object,
with very similar stellar parameters, falls, like HD\,46769, just outside 
the currently known instability domains for gravity modes discovered in
evolved (post-main sequence) stars (Lefever et al. \cite{Lefever}, Saio
et al. \cite{Saio}). This could mean that in late-type B supergiants other 
yet unknown oscillation modes with periods of a few hours might play a 
non-negligable role as well.

\section{Conclusions}\label{summary}

We reported on the discovery of a 1.59\,h period in the Galactic late-type B 
supergiant star HD\,202850. This period was found based on time-series
of optical spectroscopic observations obtained on four different nights
distributed over a time interval of 19 months. Our data
reveal simultaneous line profile variations in photospheric lines of
\ion{Si}{ii}\,$\lambda\lambda 6347, 6371$ and \ion{He}{i}\,$\lambda 6678$.
A moment analysis of these lines shows that all lines vary simultaneously 
in both the radial velocity and the skewness of the line profile, while the 
equivalent widths of the lines do not change noticeably over the time span of the 
observations. The second moment is too noisy to allow a proper analysis,
but a variation according to an axisymmetric mode might be present. 
The interpretation of these results excludes the scenario of a chemically 
peculiar star with surface abundance inhomogeneities in the form of spots. 
Instead, the results speak in favour of pulsations as a plausible origin of 
the observed variations. While non-radial pulsations in B supergiants have 
been identified previously as due to gravity modes with periods on the order 
of days, our discovery of the second late-type B supergiant with a period on 
the order of hours might open new perspectives in the studies of 
asteroseismology in both observations and theory. We hence intend 
to persue higher resolution spectroscopic time-series to determine the 
underlying physical process that causes the short-term variations discovered 
in the late B-type supergiant HD\,202850.


\begin{acknowledgements}
We thank the anonymous referee for the valuable comments on the manuscript.
This research made use of the NASA Astrophysics Data System (ADS).
M.K. and M.E.O. acknowledge financial support from GA\v{C}R under grant number
P209/11/1198. The Astronomical Institute Ond\v{r}ejov is supported by the 
project RVO:67985815.
\end{acknowledgements}



\end{document}